\documentclass[10pt,pra,letterpaper,twocolumn,showpacs,amssymb]{revtex4}

\usepackage{graphics}
\usepackage{amsfonts}
\usepackage{mathrsfs}

\def\bra#1{\mathinner{\langle{#1}|}}
\def\ket#1{\mathinner{|{#1}\rangle}}
\def\braket#1{\mathinner{\langle{#1}\rangle}}

{\catcode`\|=\active 
  \gdef\Braket#1{\begingroup
\mathcode`\|32768\let|\BraVert\left<{#1}\right>\endgroup}}
\def\BraVert{\egroup\,\mid@vertical\,\bgroup}

\begin{document}           

\title{Quantum Zeno and anti-Zeno effects in an Unstable System with Two Bound State
}

\author{Kavan Modi}
\email[Email: ]{modik@physics.utexas.edu}
\author{Anil Shaji}
\affiliation{Department of Physics, Center for Complex Quantum Systems\\
The University of Texas at Austin, Austin, Texas 78712-1081}

\date[Received]{ \today}


\begin{abstract}

We analyze the experimental observations reported by Fischer \emph{et al} [in Phys.
Rev. Lett. {\bf 87}, 040402 (2001)]  by considering a system of coupled
unstable bound quantum states $\ket{A}$ and $\ket{B}$. The state $\ket{B}$ is
coupled to a set of continuum states $\ket{C\Theta(\omega)}$.  We investigate
the time evolution of $\ket{A}$ when it decays into $\ket{C\Theta(\omega)}$ via
$\ket{B}$, and find that frequent measurements on $\ket{A}$ leads to both the
quantum Zeno effect and the anti-Zeno effects depending on the frequency of
measurements.  We show that it is the presence of $\ket{B}$ which allows for
the anti-Zeno effect.
\end{abstract}


\maketitle


\section{Introduction}

The quantum Zeno effect, first predicted by Misra and Sudarshan
\cite{bib1,PhysRevD.16.520},  is the hindrance of the time evolution of a
quantum state when frequent measurements are performed on it.  In the limit of
continuous measurement the time evolution of the state, in principle, completely
stops.  The seminal paper by Misra and Sudarshan proves the existence of an
operator corresponding to continuous measurement belonging to the Hilbert space
of a generic quantum system.  More recently, several authors have suggested that
the opposite of quantum Zeno effect may also be true
\cite{PhysRevA.56.1131,Nature.405.546,PhysRevLett.86.2699}.  That is, frequent
measurements can be used to accelerate the decay of an unstable state.  This
effect is known as the anti-Zeno effect or the inverse Zeno effect.  The
original formulation of the quantum Zeno effect treated the measurement process
as an idealized von-Neumann type; that is an instantaneous event that
induces discontinuous changes in the measured system.  The anti-Zeno effect was
first identified as a possibility when measurement processes that take a
finite amount of time were considered. This led to the suggestion
by several authors that the anti-Zeno effect should be observed more often in
physical systems than the quantum Zeno effect.

Experimental evidence supporting the quantum Zeno effect in particle physics
experiments was first pointed out by Valanju \emph{et al}
\cite{PhysRevD.21.1304,Valanju:1980mi}. Direct experimental observation of the quantum Zeno 
effect was obtained by Itano \emph{et al} \cite{PhysRevA.41.2295} in a three-level
oscillating system.  Recently, in a set of experiments Fischer,
Gutierrez-Medina, and Raizen observed, for the first time, \emph{both} the quantum Zeno
effect and the anti-Zeno effect in an {\em unstable} quantum mechanical system
\cite{PhysRevLett.87.040402}.  In this Letter we present a simple model that
reproduces all of the important results of this experiment.

This Letter is organized as follows:  We briefly describe the experiment by
Fischer \emph{et al} in section \ref{exp}.  In section \ref{mod}, we present a
simplified model of the system studied experimentally in
\cite{PhysRevLett.87.040402}. The model is exactly solvable and the time
dependence of the survival probability of the initial unstable state can be
analytically calculated.  In section \ref{num}, we show that the solutions
reproduce all the important features of the experimental system.  On the basis
of our model we argue that the anti-Zeno effect is observed because of the
presence of more than one unstable bound states in the system.  Our conclusions
are in section \ref{con}.

\section{Description of the Experiment}\label{exp}

In the experiment by Fischer \emph{et al}~\cite{PhysRevLett.87.040402}, sodium atoms
were placed in a classical magneto-optical trap that could be moved in space. 
The motional states of the atoms in the trap were studied.  Initially, the atoms
were placed in the ``ground" state of the moving trap so that they remained
inside the trap. The stable bound states occupied by the atoms in the trap are made
unstable by accelerating the atoms along with the trap at different rates.
By tuning the acceleration appropriate conditions are created for the atoms to
quantum mechanically tunnel through the barrier into the continuum of available
free-particle states. The number of atoms that tunneled out of the bound state
inside the trap as a function of time was estimated at the end of the experiment
by recording the spatial distribution of the atoms. Since the trap was
accelerated throughout the experiment, atoms that spent more time in the and
trap had higher velocities and they moved farther in unit time. So by
taking a snap-shot of the spatial distribution of all the atoms at the end of
the experiment the time at which each one of them tunneled out of the trap
can be estimated.

\begin{figure}[t]
\resizebox{4.2 cm}{3.7 cm}{\includegraphics{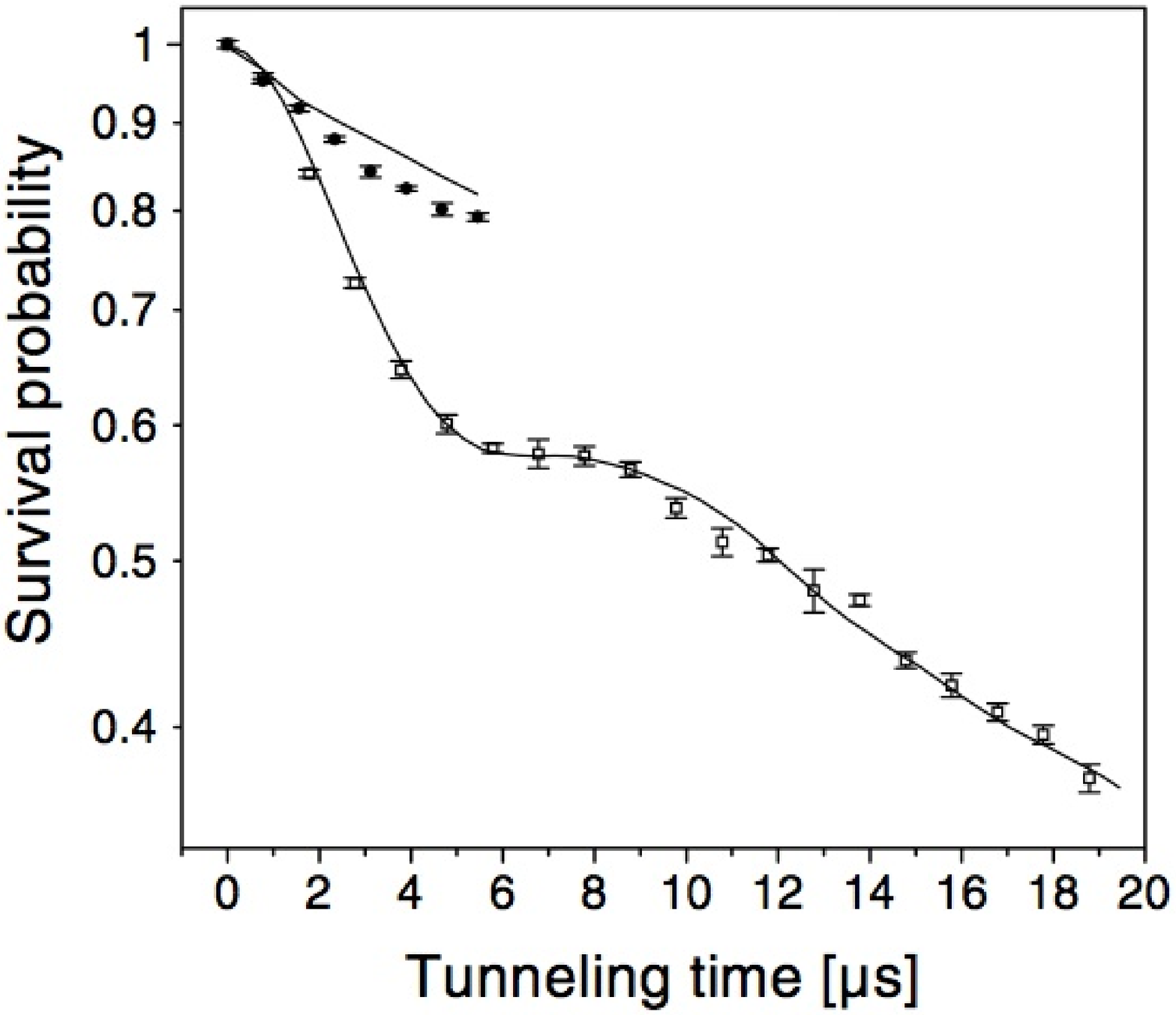}}
\resizebox{4.2 cm}{3.7 cm}{\includegraphics{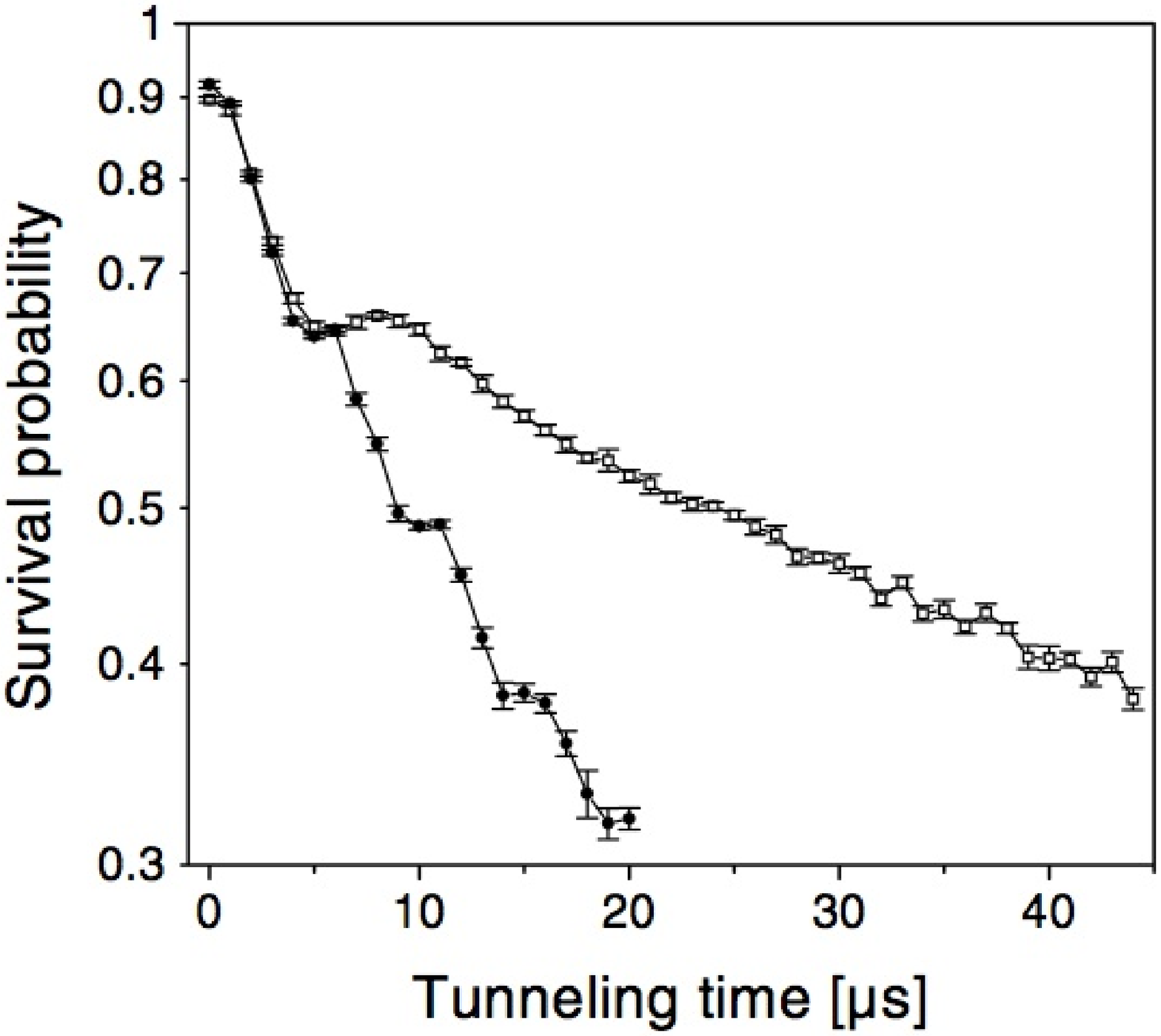}}
\caption{On the left, the lower line is the ``unmeasured" decay curve
corresponding to the case where the trapped atoms are accelerated with no
interruptions so that tunneling out of the trap is always present. The
upper line corresponds to the case where the tunneling is interrupted every
1 $\mu$s leading to the quantum Zeno effect. On the right, the upper line is 
the ``unmeasured" decay and the lower line corresponds to interruptions
every 5 $\mu$s leading to the anti-Zeno effect in the experiment by Fischer et
al.\label{sur}}
\end{figure}

To obtain the Zeno and the anti-Zeno effects, the tunneling of the atoms out of
the trap has to be interrupted by a measurement that estimates the number of
atoms still inside the trap. Such a measurement was implemented in the
experiment by abruptly changing the acceleration of the trap so
that tunneling from the ground state is temporarily halted. These interruption
periods were long enough ($40\mu s$) to separate out the atoms that tunnel out
before and after each interruption into resolvable groups. By measuring the
number of atoms in each group and knowing the total number of atoms that were
initially in the trap, the number of atoms in the trap at the beginning of each
interruption period was estimated. Using this data, the time dependence of the
survival probability of the bound motional states of an atom was reconstructed.
Their observations are summarized in Fig. \ref{sur}. 

The frequency with which interruption periods were applied determined whether
the Zeno or anti-Zeno effects were obtained. Repeatedly interrupting the system
once every micro-second led to the quantum Zeno effect. With an interruption
rate of $5\mu s$ the anti-Zeno effect was obtained.

In Fig\ref{sur}, the zero slope for the survival probability at $t=0$ is expected for the time
evolution of a generic unstable quantum state on the basis of the analyticity of
the survival probability and its time reversal symmetry 
\cite{PhysRev.123.1503}. This non-exponential behavior at short times
leads to the quantum Zeno effect.   

The shape of survival probability as a function of time of the ``unmeasured
system" has an inflection point at $t\approx7\mu s$. (see Fig. \ref{sur}).   We
show that it is the presence of a second unstable bound state 
that is responsible for this inflection point.  The anti-Zeno effect is obtained
by taking advantage of that inflection point. 

In an earlier experiment that led to this experiment, Bharucha et
al. observed tunneling of sodium atoms from an accelerated trap
\cite{PhysRevA.55.R857}.  Our analysis of the experiment in
\cite{PhysRevLett.87.040402} is motivated by the
following passage from \cite{PhysRevA.55.R857}:

\begin{quotation}
When the standing wave is accelerated, the wave number changes in time and the
atoms undergo Bloch oscillations across the first Brillouin zone. As the atoms
approach the band gap, they can make Landau-Zener transitions to the next band.
Once the atoms are in the second band, they rapidly undergo transitions to the
higher bands and are effectively free particles.
\end{quotation}

The last sentence above suggests that higher energy bound states are present in their system. An atom in the ground state might have to go through these intermediate bound states before it can tunnel out to the continuum of available free-particle states.  We investigate the effect of these intermediate states on survival probability of the ground state.


\section{\label{model}Model}\label{mod}

We consider an interacting field theory of four fields labeled $A$, $B$,
$C$, and $\Theta$, with the following commutation relations:
\begin{eqnarray*}
[a,a^{\dagger}]=[b,b^{\dagger}]=[c,c^{\dagger}]=1, \vphantom{\bigg[}\\
\left[\theta(\omega),\theta^{\dagger}(\omega'))\right]=\delta(\omega-\omega').
\end{eqnarray*}
All other commutators being zero.  $a^{\dagger}$ ($a$) etc. represent the
creation (annihilation) operators corresponding to the four fields.  Only $\Theta$ is
labeled by a continuous index $\omega$, while other fields are assumed to only
have discrete modes.  The allowed processes in the model are 
\[ A\leftrightharpoons B \; \text{and} \;  B\leftrightharpoons C\Theta. \]

The Hamiltonian for the model with these allowed processes can be written down
as
\begin{eqnarray}
\label{h0}
H=H_0+V
\end{eqnarray}
where,
\begin{eqnarray}
\label{h1}
H_0=E_A a^{\dagger} a +E_B b^{\dagger} b
+\int^{\infty}_{0}d\omega\;\omega\;\theta^{\dagger} (\omega) \theta(\omega) 
\end{eqnarray}
and
\begin{eqnarray}
\label{h2}
V&=&\Omega \; a^{\dagger}\; b +\Omega^{*}\; b^{\dagger}\; a\nonumber\\
&+&\int^{\infty}_{0}d\omega \left[f(\omega)\; b^{\dagger}\; c\; \theta(\omega)+
f(\omega)^{*}\; c^{\dagger} \theta^{\dagger}(\omega)\; b \right].
\end{eqnarray}

\begin{figure}[t]\label{eng}
\resizebox{8 cm}{5 cm}{\includegraphics{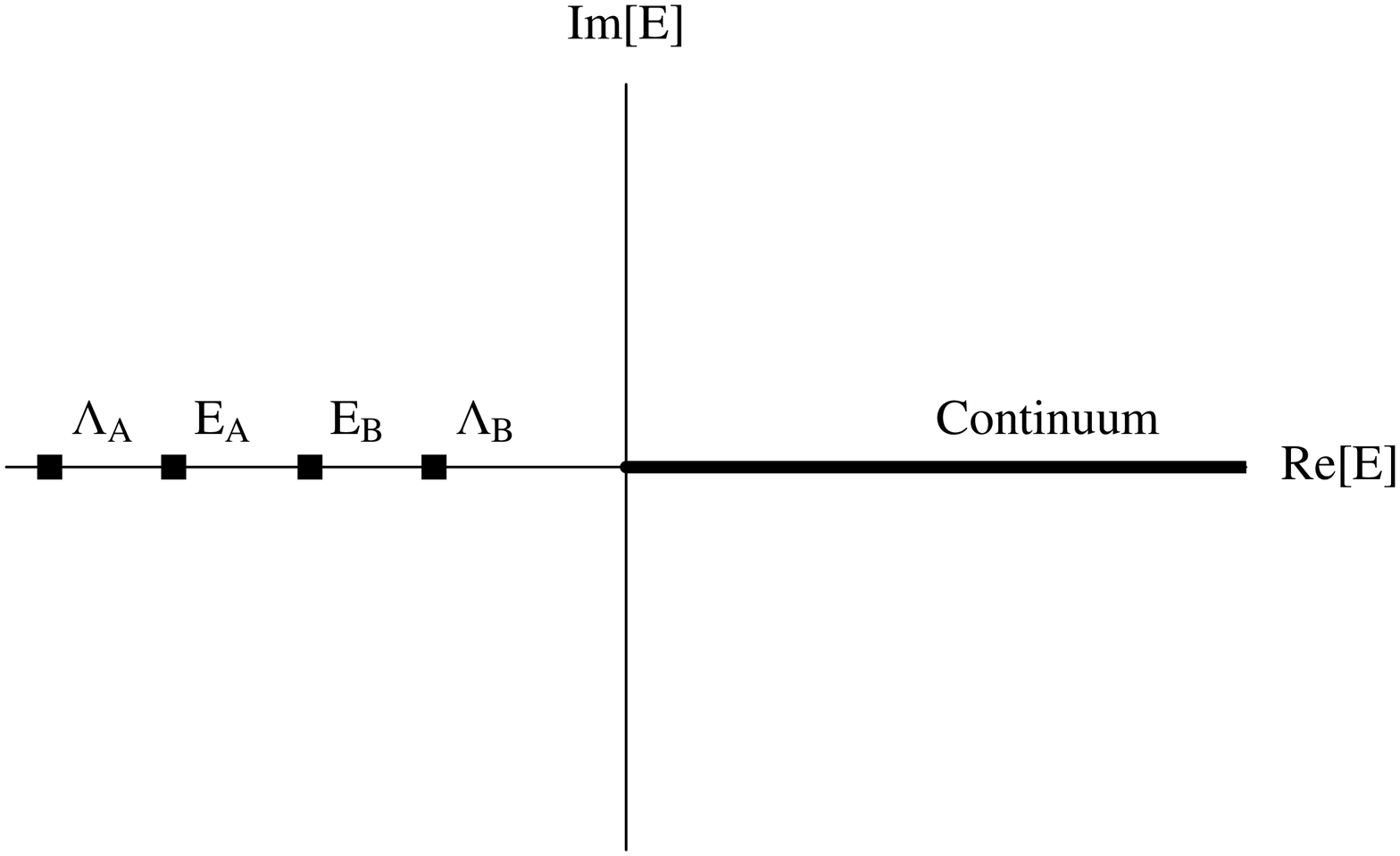}}
\caption{$E_{A,B}$ are the energies of bare bound states. We choose $E_A$ and
 $E_B$ to be real and negative. The full Hamiltonian, $H$, then 
has two real negative eigenvalues indicated by $\Lambda_{A}$ and $\Lambda_B$.
$H$ also has a continuum of eigenstates along the
positive real-axis.\label{eng1}}
\resizebox{8 cm}{5 cm}{\includegraphics{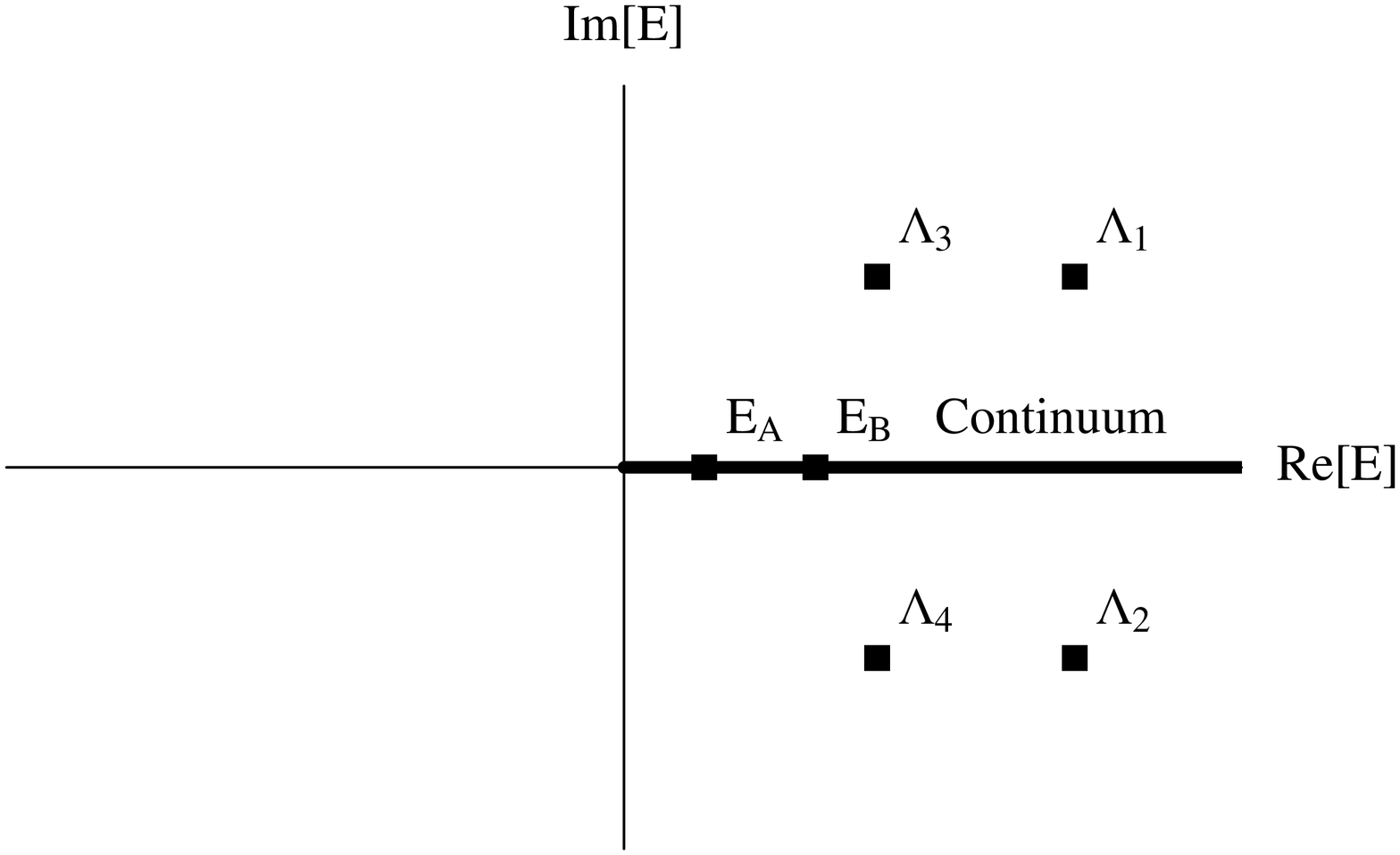}}
\caption{The discrete eigenvalues of $H_0$, $E_{A,B}$ are chosen to be
real and positive. The full Hamiltonian has no real negative eigenvalues
corresponding to bound states. The resolvent of $H$ has two complex poles. The
two poles represent unstable states and they could have either positive or
negative imaginary parts. These four possibilities are indicated by
$\Lambda_{1,2,3,4}$. The sign of the imaginary part is fixed by the boundary
conditions. $\Lambda_2$ and $\Lambda_4$ correspond to unstable states that
decay in time. The continuum of physical states lies along positive
real-axis.\label{eng2}}
\end{figure}

The two discrete energy levels are denoted by $E_A$ and $E_B$. The Hamiltonian
in Eq.~(\ref{h0}) is obtained by modifying the Hamiltonian for the
Friedrichs-Lee model \cite{bib11,PhysRev.95.1329}. It is instructive to look
at the spectrum of $H$ and $H_0$ in the complex plane at this point. We call
the eigenstates of $H$ physical states, while the eigenstates of $H_0$ are
referred to as bare states. We choose the zero point of energy so
that the continuum eigenstates of $H_0$ have positive energy ($0 \leq \omega
< \infty$). If $E_A$ and $E_B$ are negative and if the shift in these
energies due to the perturbation $V$ is small then the spectrum of the physical
Hamiltonian, $H$, will contain two stable bound states with negative energies
$\Lambda_A$ and $\Lambda_B$ in addition to continuum of states with positive
energies. This is illustrated in Fig. \ref{eng1}.  

We are interested in studying the temporal evolution of an unstable system. So we
choose $E_A$ and $E_B$ to be positive so that they lie embedded in the physical continuum
spectrum. The spectrum of the physical Hamiltonian will no longer include bound
states. The eigenstates of $H$ belonging to the continuum, corresponding to
eigenvalues $0\leq\lambda<\infty$, will form a complete set of states. To see what
happened to the two bound states of the bare Hamiltonian $H_0$ when the
perturbation $V$ is introduced, it is instructive to look at the resolvent of
the physical Hamiltonian, $(E-H)^{-1}$. The resolvent, in this case, will have
two complex poles indicating two transient or unstable states. The location of
these poles is fixed by choosing appropriate boundary conditions to make sure
that the unstable states decay (rather than grow) in time. This is illustrated
in Fig. \ref{eng2}.

\subsection{The Continuum States}

We are interested in the time evolution of the eigenstates of $H_0$, namely the
two bound bare states $\ket{A}$ and $\ket{B}$, and the continuum states
$\ket{C\Theta(\omega)}$. The state $\ket{A}$ in our model corresponds to the
unstable bound state occupied by the atoms inside the trap in the experiment by
Fischer \emph{et al} The states $\ket{C\Theta(\omega)}$ represent the continuum
outside the trap into which the bound state can decay.

We have introduced an additional unstable bound state $\ket{B}$  which
represents a second bound motional state of the trap. The state $\ket{A}$ is
directly coupled to only $\ket{B}$ and the decay of $\ket{A}$ into
$\ket{C\Theta}$ is mediated by the new state $\ket{B}$. We will show that
the presence of the additional bound state can explain several of the key
features of the experiment by Fischer \emph{et al}

We begin by writing the full Hamiltonian, $H$, in matrix form using eigenstates
of $H_{0}$ as basis \cite{bib10},
\begin{equation}\label{e1}
H=\left(\matrix{E_A&\Omega^{*}&0\cr
\Omega&E_B&f^{*}(\omega')\cr
0&f(\omega)&\omega  \delta (\omega -\omega')\cr}\right).
\end{equation}
Let $\ket{\psi_\lambda}$ represent an eigenstate of $H$ with eigenvalue
$\lambda$, satisfying the eigenvalue equation
\begin{equation}\label{eigeq}
H\psi_{\lambda}=\lambda\psi_{\lambda}.
\end{equation}
We express $\psi_\lambda$ also in terms the eigenstates of $H_0$,
\begin{equation}\label{e2}
\psi_{\lambda}=\left(\matrix{
\braket{A|\psi_\lambda}\cr
\braket{B|\psi_\lambda}\cr
\braket{C\Theta(\omega)|\psi_\lambda}\cr}\right)\equiv
\left(\matrix{\mu^{A}_{\lambda}\cr\mu^{B}_{\lambda}\cr\phi_{\lambda}(\omega)\cr}
\right).
\end{equation}
$H$ can have three possible classes of eigenstates; a maximum of two bound
states and a set of continuum eigenstates.  We first look at the continuum
eigenstates  of $H$ followed by the remaining physical bound states in next
section.  

Using the Eq.~(\ref{eigeq}), we get a system of three coupled integral
equations:
\begin{eqnarray}\label{e3}
\mu^{A}_{\lambda}&=& {\Omega^{*}\over\lambda-E_A}\mu^{B}_{\lambda} \;\;\;\;\;
(\lambda\neq E_A),\\\label{e5}
\mu^{B}_{\lambda}&=&{{\Omega\mu^{A}_{\lambda}
+\int_{0}^{\infty}{d\omega'}{f^{*}(\omega') \phi_{\lambda}(\omega')}}\over{\lambda-E_B}}\;\;\;\;\;
(\lambda\neq E_B),\\\label{e4}
\phi_{\lambda}&=&{f(\omega)\over\lambda-\omega+i\epsilon}\mu^{B}_{\lambda}
+\delta(\lambda-\omega).
\end{eqnarray}
The continuum wave function in Eq.~(\ref{e4}) is singular in the energy
(momentum) space, therefore it extends to infinity in the configuration space.
We choose the ``in" solution by choosing the sign of the imaginary part. 
Eqs.~(\ref{e3}-\ref{e4}) can be solved simultaneously to get
\begin{eqnarray}\label{e6}
\psi_{\lambda}&=& \left(\matrix{
\mu^{A}_{\lambda_{\;}} \cr
\mu^{B}_{\lambda_{\;}} \cr
\phi_{\lambda}(\omega) \cr} \right) 
= \left(\matrix{
{f(\lambda)\over\beta^{+}(\lambda)}{\Omega^{*}\over{\lambda-E_A}}\cr
{f(\lambda)\over\beta^{+}(\lambda)} \cr
{f(\lambda)\over\beta^{+}(\lambda)}{f(\omega)\over{\lambda-\omega+i\epsilon}}+
\delta(\lambda-\omega)\cr} \right)
\end{eqnarray}
where
\begin{eqnarray}\label{e7}
\beta(z)=z-E_B-{\Omega^{2}\over{z-E_A}}
-{\int_{0}^{\infty} 
{{|f(\omega)|^{2}\over{z-\omega}}d\omega}}, 
\end{eqnarray}
is a real analytic function and
\begin{eqnarray}\label{e8}
\beta^{\pm}(\lambda)=\beta(\lambda\pm i\epsilon)   \;\;\;\;\text{and}\;\;\;\;
{\beta^{\pm}(\lambda)}^{*}=\beta^{\mp}(\lambda).
\end{eqnarray}


\subsection{The Physical Bound States}

Now we look at the case when $\lambda\neq\omega$ for all $\lambda$.  Since
$0\leq\omega<\infty$, the physical states in this case are restricted to
eigenvalues on the negative real-axis.  The delta function is no longer present
in Eq.~(\ref{e4}), meaning that the physical states do not extend to infinity. 
Such solutions of the eigenvalue problem correspond to the stationary
eigenstates of $H$.  Eq.~(\ref{e4}) now becomes

\begin{equation}\label{e9}
\phi_{\lambda}={f(\omega)\over\lambda-\omega}\mu^{B}_{\lambda}.
\end{equation}
Solving for $\mu^{B}_{\lambda}$, we get
\begin{equation}\label{e10}
\beta(\lambda)\mu^{B}_{\lambda}=0.
\end{equation}
We choose $\mu^{B}_{\lambda}\neq0$, to obtain a non-trivial solution, then
$\beta(\lambda)=0$, which leads to the following expression, with zeroes at
$\lambda=\Lambda_{j}$.
\begin{eqnarray}\label{quad}
\lambda^{2}-
\lambda\left(E_A+E_B+{\int_{0}^{\infty} 
{{|f(\omega)|^{2}\over{\lambda-\omega}}d\omega}}\right) \hspace{8 mm}
\nonumber \\
 -\Omega^{2} +\left. E_A\left(E_B+{\int_{0}^{\infty} 
{{|f(\omega)|^{2}\over{\lambda-\omega}}d\omega}}\right)\right|_{\Lambda_{j}}&
\!\!\!\!\!\!=0
\end{eqnarray}

The wave functions belonging to this class of solutions are just complex numbers
that are fixed by the normalization condition. The solutions are 
the following:
\begin{equation}
\psi_{\Lambda_{j}}= \left(\matrix{
\mu^{A}_{\Lambda_{j}} \cr
\mu^{B}_{\Lambda_{j}} \cr
\phi_{\Lambda_{j}} \cr} \right)=\left(\matrix{
{1\over\sqrt{\beta'(\Lambda_{j})}}{\Omega^{*}\over{\Lambda_{j}-E_A}}\cr
{1\over\sqrt{\beta'(\Lambda_{j})}} \cr
{1\over\sqrt{\beta'(\Lambda_{j})}}{f(\omega)\over{\Lambda_{j}-\omega}}
\cr} \right),
\end{equation}
where
\begin{eqnarray}\label{e33}
\beta'(\Lambda_{j})\!\!\!&= &\!\!\!\left.{d\beta\over{d\lambda}}\right|_{
\lambda=\Lambda_{j}} \\
&=&\!\!\!
\left.1+{|\Omega|^2\over{(\lambda-E_A)^2}}\!+\!\!{\int_{0}^{\infty} \!\!{
|f(\omega)|^2\over{(\lambda-\omega)^2}}d\omega}\right|_{\lambda=\Lambda_{j}}
.\nonumber
\end{eqnarray}

We can find $\Lambda_j$ by numerically solving  Eq.~(\ref{quad}).  For a weak
potential $V$, we can roughly approximate
\begin{eqnarray}\label{la}
\Lambda_j\approx E_{A,B} \pm i \, {\mbox{Im}}\left[ \left. \int_{0}^{\infty}
{{\left|f(\omega)\right|^2}\over{\lambda-\omega+i\epsilon}}
\right|_{\lambda=E_{A,B}}\right].
\end{eqnarray}

In general,  Eq.~(\ref{quad}) has four possible roots, given by  Eq.~(\ref{la}).
Only the bound states with real eigenvalues $\Lambda_j$ are physically relevant.
 Once again this due to the fact that the physical spectrum is confined to the
real-axis.  The imaginary part of the integrals in  Eq.~(\ref{la}) vanishes
everywhere except near $\omega_0$. Hence for $E_A$ and $E_B$ with
sufficiently negative energies there are two physical bound states with real
eigenvalues.  Once again this will not be the case of interest since no decay
takes place here. For sake of completeness it should be noticed that the
physical bound states will be present wherever  the imaginary part of the
integral in Eq.~(\ref{la}) vanishes.


\subsection{Survival Amplitude}

We are now in a position to calculate the survival amplitude of the bare state
as a function of time.  If $\ket{A}$ is occupied at t=0, then its survival
amplitude is obtained by computing the matrix element $\mathscr{A}_{A}(t) =
\braket{A|e^{-iHt}|A}$. We can compute this matrix element by inserting a
complete set of physical states, i.e.
\[ \int_{0}^{\infty} \ket{\psi_\lambda} \bra{\psi_\lambda}
d\lambda + \ket{\psi_{\Lambda_{A}}} \bra{\psi_{\Lambda_{A}}} +
\ket{\psi_{\Lambda_{B}}} \bra{\psi_{ \Lambda_{B}}}=1. \]
Here we are considering the possibility that stable bound states $\ket{\psi_{\Lambda_{A,B}}}$ with real energies $\Lambda_{A,B}$ may exist. Let's continuing with the calculation of the survival amplitude.
\begin{eqnarray}\label{e12}
\mathscr{A}_{A}(t) & = & \int_{0}^{\infty} \braket{A|e^{-iHt}|\psi_\lambda}
\braket{\psi_\lambda|A} d\lambda \nonumber \\
&& \hspace{10 mm} + \braket{A|e^{-iHt}|\psi_{\Lambda_{A}}}
\braket{\psi_{\Lambda_{A}}|A}  \nonumber \\
&& \hspace{20 mm} +\braket{A|e^{-iHt}|\psi_{\Lambda_{B}}}
\braket{\psi_{\Lambda_{B}}|A } \vphantom{\bigg[} \nonumber\\
&=&\int_{0}^{\infty} e^{-i\lambda t} \braket{A|\psi_{\lambda}}
\braket{\psi_{\lambda}|A} d\lambda \nonumber \\
&& \hspace{10 mm} + e^{-i \Lambda_{A} t} \braket{A|\psi_{\Lambda_{A}}}
\braket{\psi_{\Lambda_{A}}|A} \nonumber \\
&& \hspace{20 mm} +e^{-i \Lambda_{B} t} \braket{A|\psi_{\Lambda_{B}}}
\braket{\psi_{\Lambda_{B}}|A} \vphantom{\bigg[} \nonumber \\
&=&\int_{0}^{\infty}e^{-i\lambda t} \left|\mu^{A}_{\lambda}
\right|^{2} d\lambda + e^{-i\Lambda_{A} t} \left|\mu^{A}_{\Lambda_{A}}
\right|^{2} \nonumber\\
&&  \hspace{33 mm} + e^{-i\Lambda_{B} t} \left|\mu^{A}_{\Lambda_{B}}
\right|^{2}.
\end{eqnarray}
We have used the completeness and orhtonormality of the physical
states in the above equation. Orthonormality of these states is
shown in Appendic A, while completeness can be demonstrated by using the
techniques described in Appendix A and in reference \cite{bib10}.

The survival probability of $\ket{A}$ is simply the square of the amplitude,
\begin{eqnarray}\label{e14}
P_{A}(t) &= & |\mathscr{A}_{A}(t)|^{2} \nonumber \\
& =& \bigg|\int_{0}^{\infty} e^{-i\lambda t} \left|\mu^{A}_{\lambda}
\right|^{2} d\lambda + e^{-i\Lambda_{A} t} \left|\mu^{A}_{\Lambda_{A}}
\right|^{2} \nonumber \\ 
&& \hspace{30 mm} + e^{-i\Lambda_{B} t} \left|\mu^{A}_{\Lambda_{B}}
\right|^{2} \bigg|^2.
\end{eqnarray}
Similarly, if state $\ket{B}$ is occupied at t=0, then the survival
probability of $\ket{B}$ will take the form
\begin{eqnarray}\label{e15}
P_{B}(t)& =& |\mathscr{A}_{B}(t)|^{2}
\nonumber \\ 
&=& \bigg| \int_{0}^{\infty}e^{-i\lambda t} \left|\mu^{B}_{\lambda}
\right|^{2} d \lambda + e^{-i\Lambda_{A} t} |\mu^{B}_{\Lambda_{A}}|^{2}
\nonumber \\
&& \hspace{30 mm} + e^{-i\Lambda_{B} t} |\mu^{B}_{\Lambda_{B}}|^{2}
\bigg|^2.
\end{eqnarray}


\section{Numerical Calculations and Results}\label{num}

In this section, we discuss the numerical calculations of the survival
probabilities $P_{A}(t)$ and $P_{B}(t)$. We choose the form factor to be
\begin{equation}\label{e16}
f(\omega)={{\sigma \mu^{2}
\sqrt{\omega}}\over{(\omega-\omega_{0})^{2}+\mu^{2}}}.
\end{equation}
The factor of $\sqrt{\omega}$ in the numerator is a phase-space contribution.
The rest of $f(\omega)$ is just the Lorentzian line shape. The
function $f(\omega)$ peaks near $\omega_0$ and its width is controlled by $\mu$,
and its strength by $\sigma$.

We require the strength of the perturbation to be weak relative to the eigenvalues
of $H_0$. In other words, we don't want the original system to change
drastically. So we choose small values for the parameters $\mu$, $\sigma$ and
$\Omega$ and set $\mu=0.30$, $\sigma=0.11$, and $\Omega=0.04$. We also want the
bare bound states to become unstable in the presence of the perturbation $V$.
This can be achieved by setting their energies to lie in the physical continuum
sufficiently above the threshold. The physical continuum ranges from zero to
infinity, and so we choose numerical values $E_A=2.00$ and $E_B=2.10$.

For the choice of $E_A$ and $E_B$ here,  Eq.~(\ref{quad}) yields complex
eigenvalues for the physical bound states.  The condition on the physical bound
states' stability requires that they have real-negative eigenvalues and so here
they are unstable and show up only as a spectral density. Since the physical
spectrum is confined to the real-axis and since there are no stable physical
bound states, the last two terms in the Eqs.~(\ref{e14}) and (\ref{e15}) are
zero. The physical continuum states $\ket{\psi_{\lambda}}$ form a complete set
of states by themselves. The equations for survival probability then reduce to
\begin{equation}\label{e17}
P_{A}(t) =\left| \int_{0}^{\infty} e^{-i\lambda t} \left|\mu^{A}_{\lambda}
\right|^{2} d \lambda \right|^{2}
\end{equation}
and
\begin{equation}\label{e18}
P_{B}(t) =\left| \int_{0}^{\infty} e^{-i\lambda t}
\left| \mu^{B}_{\lambda}\right|^{2} d \lambda \right|^{2}.
\end{equation}


\subsection{``Unmeasured" Evolution}

\begin{figure}[!t]
\resizebox{8.1 cm}{5.1 cm}{\includegraphics{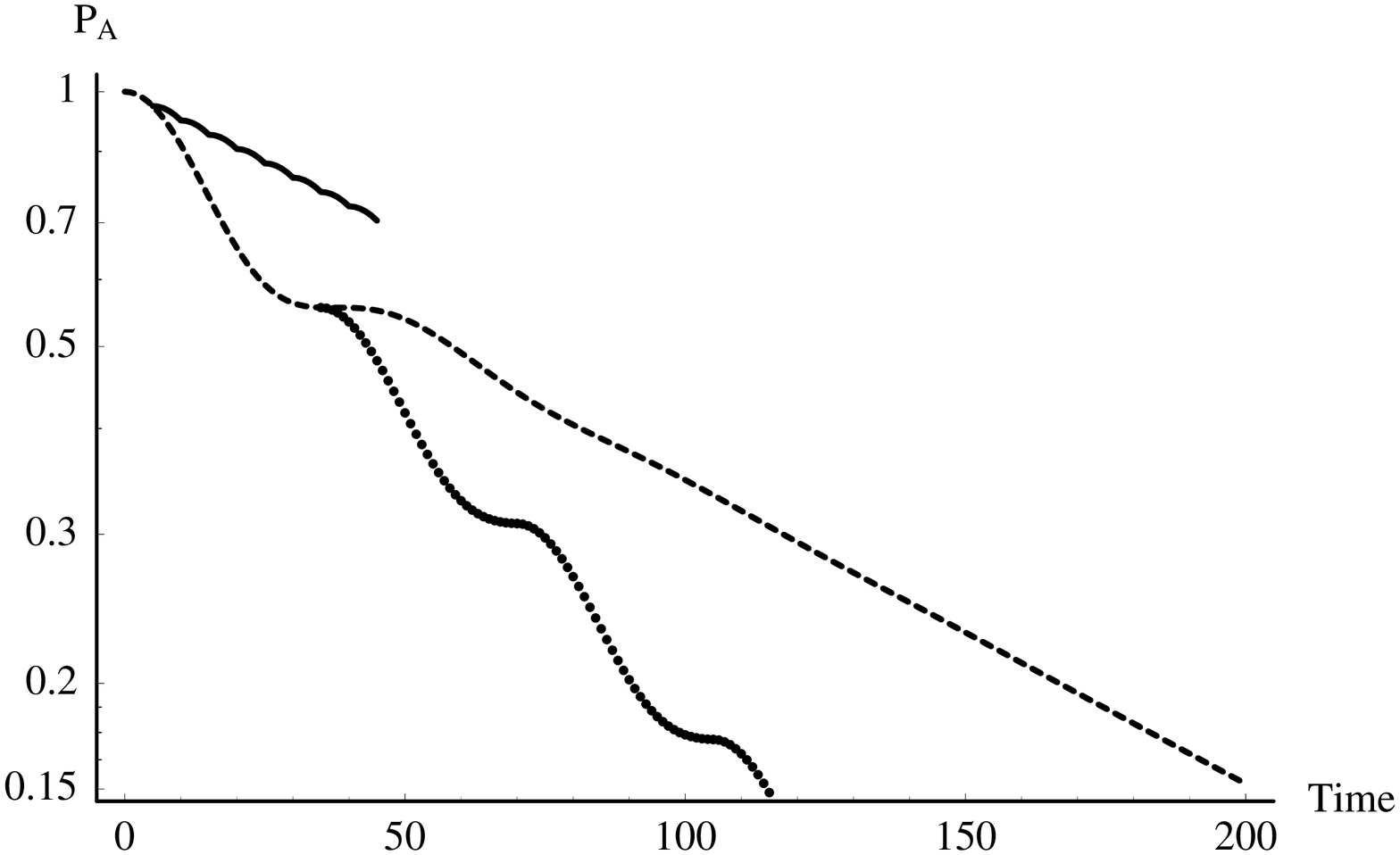}}
\caption{Survival probability of $\ket{A}$ given by Eq.~(\ref{e17}). The dashed
line shows the ``unmeasured" evolution of $\ket{A}$. The solid line shows the
effect of repeated measurements made at high frequencies leading to the
Zeno effect. The dotted line show the effect of repeated measurements made at
lower frequencies leading to the anti-Zeno effect. The parameter values used are
$E_A=2.00$, $E_B=2.10$, $\omega_0=2.10$, $\mu=0.30$, $\sigma=0.11$, and
$\Omega=0.04$.\label{zenoA}}

\resizebox{8.1 cm}{5.1 cm}{\includegraphics{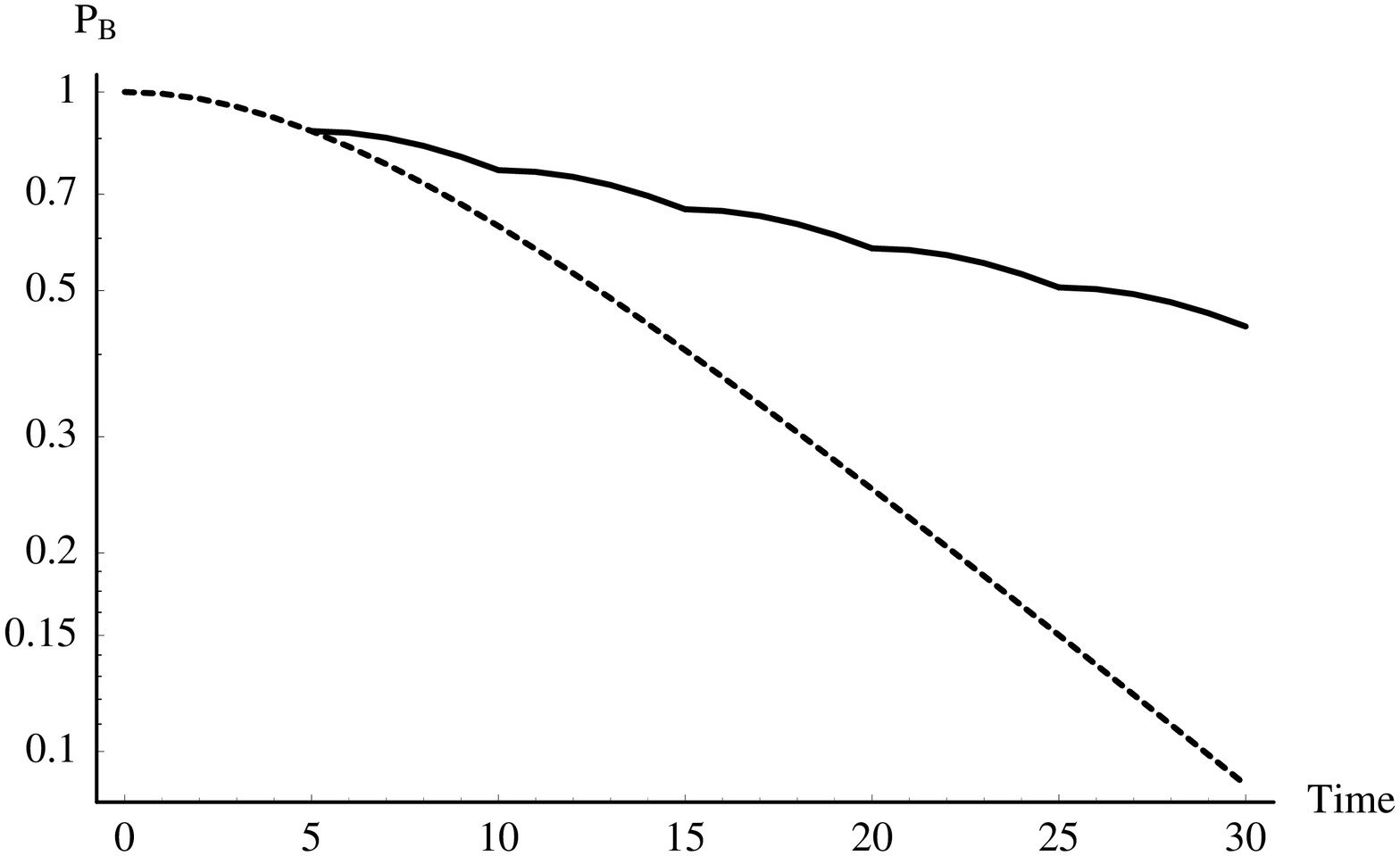}}
\caption{Survival probability of $\ket{B}$ given by  Eq.~(\ref{e18}) with
$(\Omega=0)$.  The dashed line shows the ``unmeasured" evolution of $\ket{B}$.
The solid line shows the quantum Zeno effect appearing due to repeated
measurements. The parameter values that were used are $E_B=2.10$,
$\omega_0=2.10$, $\mu=0.30$, $\sigma=0.11$, and $\Omega=0$.
\label{zenoB}}
\end{figure}

First, we study the survival probability as a function of time of $\ket{A}$
($P_A$). We then compare it to the survival probability of $\ket{B}$ ($P_B$)
with $\Omega\rightarrow0$.   In the second case the first bound state $\ket{A}$ is
completely cutoff from the rest of the system as seen from  Eq.~(\ref{e1}). Then
 we are dealing with a system with only one bound state coupled
to the continuum.  The differences between these two cases show precisely the
effects that the second bound state has on the survival probability of
$\ket{A}$. Note that the second case is nothing more than the well known
Fredrichs-Lee Model.  The form factor $f(\omega)$ is peaked
around the $E_B$ and so we have $\omega_0=2.10$. The ``unmeasured"
survival probability of the states $\ket{A}$ and $\ket{B}$ are the dashed curves
in Figs. \ref{zenoA} and \ref{zenoB} respectively.  Notice that in both cases
for long times the decay is exponential.

\subsection{Effective Evolution: Zeno and anti-Zeno}

Interrupting the system by changing the acceleration is a necessary step in the
experiment by Fischer \emph{et al} to create the quantum Zeno effect or the anti-Zeno
effect.  The interruption resets the system and the tunneling process has to
restart after the interruption period. In our model the shifting of $\Omega$ is
analogous to changing the acceleration in the experiment.  The time evolution of
the system (when $\Omega=0.04$) can be hindered by shifting $\Omega$ to a value
much smaller than the difference between $E_A$ and $E_B$. This effectively cuts
off the oscillations between $\ket{A}$ and $\ket{B}$. During the interruptions,
$\ket{B}$ is still connected to the continuum, just as in the experiment.  The
population of $\ket{A}$ remains constant during the interruption periods.
The length of the interruption period is important for resolving the different
momentum states in the experiment by Fisher \emph{et al} as shown in the last figure
in \cite{PhysRevLett.87.040402}. We do not have this constraint in our
model and our measurements can treated as instantaneous (von-Neumann type).   

The experimental results presented in \cite{PhysRevLett.87.040402} show the
effective evolution of the atoms in the trap with the interruption periods
removed. We also look at the effective evolution, with the interruption periods
 filtered out. Numerically, the procedure for obtaining the time evolution
with interruptions is simple. We start by fixing the interval of time, $\tau$,
between the measurement induced interruptions. We start with a bare state with
unit amplitude and compute its survival porbability till time $\tau$. At this
point the measurement is assumed to reset the system. The initial bare state
wave function had only one non-zero component when expressed in the basis of
bare states. Time evolution of this state under the full Hamiltonian makes all
three components non-zero in general. Resetting the system correponds to setting
the two new components that appeared as a result of the evolution back to zero.
This new (un-normalized) state is the starting point for further evolution until
the next interruption. This process is repeated several times to obtain the
graph of the survival probability of the initial unstable state when it is
subject to frequent interruptions.  

The solid line and the dotted line in Fig.~\ref{zenoA} shows how the effective
time evolution of $\ket{A}$ can be hindered or accelerated by repeatedly
interrupting the system. The second inflection point in the unmeasured evolution
of the state $|A\rangle$ that allows us to obtain the anti-Zeno effect is
present due to the second unstable state $|B \rangle$. In the second scenario when
we have set $\Omega=0$ there is only one unstable state in the system. From the
graph of the survival probabilty of the unstable state without interruptions
shown by the dashed line in Fig. \ref{zenoB} we see that there is no way we can
choose an interruption frequency that will lead to the anti-Zeno effect. On the
other hand, interruptions made very frequently can lead to the quantum Zeno
effect as shown by the solid line in Fig.~\ref{zenoB}.


\section{Conclusion}\label{con}

We present a simple theoretical model of the experiment done by Fischer \emph{et al} 
In our model state $\ket{A}$ represents the motional ground state in their experiment.
$\ket{B}$ represents a higher energy motional bound state, while $\ket{C\Theta(\omega)}$ 
represents the continuum of available free-particle states in their experiment. The presence of
the intermediate states is clearly stated in reference \cite{PhysRevA.55.R857}. 
We study the survival probability of the ground state as it decays into the 
set of continuum states via an intermediate bound state. Our results are in agreement with the results of the experiment.  

We have shown how repeated interruptions of the time evolution of an 
unstable state can lead to the Zeno effect in our model. The anti-Zeno 
effect is also obtained in a similar fashion but we find that the system must 
have additional peculiarities for obtaining this effect. In our model there is 
an additional unstable state that mediates the decay of the original unstable 
state. The presence of this new state is shown to produce an inflection point 
in the graph of the survival probability of the unstable state that is of interest. 
This inflection point defines a time scale at which repeated measurements 
may be done on the system to effectively speed up its decay and thus obtain 
the anti-Zeno effect. We also point out that the second bound state is not 
needed to obtain the Zeno effect. A generic decaying system can exhibit the 
Zeno effect if its time evolution is very frequently interrupted by measurements. 
We note here that a similar analysis has been done for oscillating systems by Panov
\cite{bib13} and on the Friedrichs Lee model by Antoniou \emph{et al} in \cite{antoniou01}.

\begin{acknowledgments}
We thank Prof. E.C.G. Sudarshan for his support in this project.  One of us
(K.M.) thanks Shawn Rice and Laura Speck for proof reading the manuscript.  A.S.
acknowledges the support of US Navy - Office of Naval research through grant
Nos.~N00014-04-1-0336 and N00014-03-1-0639.
\end{acknowledgments}


\appendix
\section{Orthonormality}

Here we want to show that the continuum states are orthonormal.  That is 
\begin{eqnarray}\label{a.1}
\psi_{\eta}^{*}\psi_{\lambda}&=&
\left(\matrix{{\mu^{A}_{\eta}}^{*} \cr {\mu^{B}_{\eta}}^{*} \cr \phi^{*}_{\eta}(\omega)
\cr}\right)
\left(\mu^{A}_{\lambda}, \mu^{B}_{\lambda}, \phi_{\lambda}(\omega)\right)\nonumber\\
\nonumber\\
&=&{\mu^{A}_{\eta}}^{*}\mu^{A}_{\lambda}+
{\mu^{B}_{\eta}}^{*}\mu^{B}_{\lambda}+
{\phi_{\eta}}^{*}(\omega)\phi_{\lambda}(\omega)\\
&=&\delta(\lambda-\eta).\nonumber
\end{eqnarray}

We start by looking at the last term in Eq.~(\ref{a.1}).
\begin{eqnarray}\label{a.2}
\phi^{*}_{\eta}(\omega)\phi_{\lambda}(\omega)&=&
{f^{*}(\eta)f(\lambda)\over{\beta^{-}(\eta)(\eta-i\epsilon-\lambda)}}
+{f^{*}(\eta)f(\lambda)\over{\beta^{+}(\lambda)(\lambda+i\epsilon-\eta)}}\nonumber\\
&& \hspace{3mm}+{f^{*}(\eta)f(\lambda)\over{\beta^{-}(\eta)\beta^{+}(\lambda)}}\times\nonumber\\
&& \hspace{10mm}\int^{\infty}_{0}d\omega{|f(\omega)|^{2}\over{
(\lambda+i\epsilon-\omega)(\eta-i\epsilon-\omega)}}\nonumber\\
&& \hspace{20mm}+\delta(\lambda-\eta) 
\end{eqnarray}
We can break up the integral in two integrals using partial fractions:
\begin{eqnarray}\label{a.3}
{1\over{(\lambda+i\epsilon-\omega)(\eta-i\epsilon-\omega)}}= 
{1\over{\eta-\lambda-2i\epsilon}}\times\nonumber\\
\left({1\over{\lambda+i\epsilon-\omega}}-{1\over{\eta-i\epsilon-\omega}}\right).
\end{eqnarray}
Using Eq.~(\ref{a.3}), we re-write the integral term in Eq.~(\ref{a.2}) as,
\begin{eqnarray}\label{a.4}
{\int^{\infty}_{0}}d\omega
{|f(\omega)|^{2}\over{(\lambda+i\epsilon-\omega)
(\eta-i\epsilon-\omega)}}={1\over{\eta-\lambda-2i\epsilon}}\times\nonumber\\
\left(\int^{\infty}_{0}d\omega{|f(\omega)|^{2}\over{\lambda+i\epsilon-\omega}}-
\int^{\infty}_{0}d\omega{|f(\omega)|^{2}\over{\eta-i\epsilon-\omega}} \right).
\end{eqnarray}
Using Eqs.~(\ref{e7}) and (\ref{e8}), we can re-write Eq.~(\ref{a.4}) in
the following manner.
\begin{eqnarray}\label{a.5}
{1\over{\eta-\lambda-2i\epsilon}}\left(
\lambda+i\epsilon-\beta^{+}(\lambda)-E_B-{|\Omega|^{2}\over{\lambda-E_A}}\right.\nonumber\\
 \hspace{4mm}\left.-\eta+i\epsilon+\beta^{-}(\eta)+E_B+{|\Omega|^{2}\over{\eta-E_A}}\right)
\end{eqnarray}
which then reduces to
\begin{equation}\label{a.6}
-1-{|\Omega|^{2}\over{(\lambda-E_A)(\eta-E_A)}} -
{\beta^{-}(\eta)\over{\lambda+i\epsilon-\eta}} - {\beta^{+}(\lambda)\over{
\eta-i\epsilon-\lambda}}.
\end{equation}
The last two terms in Eq.~(\ref{a.6}) cancel the first two terms in
Eq.~(\ref{a.2}), and it reduces to
\begin{eqnarray}\label{a.8}
\phi^{*}_{\eta}\phi_{\lambda}=
&-&
{f^{*}(\eta)f(\lambda)\over{\beta^{-}(\eta)\beta^{+}(\lambda)}}{|\Omega|^{2}
\over{(\eta-E_A)(\lambda-E_A)}}\nonumber\\
&&\hspace{10mm}-{f^{*}(\eta)f(\lambda)\over{\beta^{-}(\eta)\beta^
{+}(\lambda)}}
+\delta(\lambda-\eta).
\end{eqnarray}
Notice, the first two terms in the last equation are the negatives of 
${\mu^{A}_{\eta}}^{*}\mu^{A}_{\lambda}+{\mu^{B}_{\eta}}^{*}\mu^{B}_{\lambda}$, and so the final result is
\begin{eqnarray}
\psi_{\eta}^{*}\psi_{\lambda}=\delta(\lambda-\eta).
\end{eqnarray}
We have now shown that the set of continuum states is orthonormal.  Using similar
techniques, we can also show that the physical eigenstates form a complete set. 

\bibliography{zeno}

\end{document}